\documentstyle[11pt,newpasp,twoside,epsf]{article}
\def\edcomment#1{\iffalse\marginpar{\raggedright\sl#1\/}\else\relax\fi}
\reversemarginpar

\begin{document}

\title{Preparing the COROT space mission: Building a photometric and
variability database of stars in its field of view}

 \author{P.J.~Amado, R. Garrido}
\affil{Instituto de Astrof\'{\i}sica de Andaluc\'{\i}a-CSIC. Camino Bajo de
 Huétor, P.O.~Box 3004. 18080. Granada. Spain}
\author{E.~Poretti}
\affil{INAF-Osservatorio Astronomico di Brera, Via E. Bianchi 46,
 23807 Merate, Italy}
\author{E.~Michel}
\affil{Observatoire de Paris, LESIA, UMR 8109, Pl. J. Jansen, 92195
 Meudon, France}

\section{Introduction}

The CNES/European space mission COROT will monitor asteroseismic
targets located in selected fields to probe stellar interiors.
Therefore, suitable candidate targets have to be searched for in order
to optimize the scientific return of the mission.  However, to be able
to use the asteroseismic tools on the stars, their physical parameters
must be known in advance.  In this work, we detail the process of
building a photometric database of all the stars brighter than $V=8.0$
mag in the field of view of COROT and the process of selecting
suitable $\delta$~Scuti and $\gamma$~Dor-type stars for the mission.

For an optimal selection of the seismology targets (for both COROT
programs devoted to asteroseismology, i.e., the core and exploratory
ones), it is essential to gather as much a priori information as
possible on all potential candidates.  To this aim,
Str\"omgren-Crawford {\em uvby}--H$\beta$ and Ca~{\sc ii}~H\&K
photometry was obtained for all of them.  These data have been used to
derive estimates of their effective temperature, surface gravity and
metallicities.  These observations, together with high resolution
echelle spectroscopy and high angular resolution imaging observations
are components of an ambitious ground-based program.

   \begin{figure}
      \plotfiddle{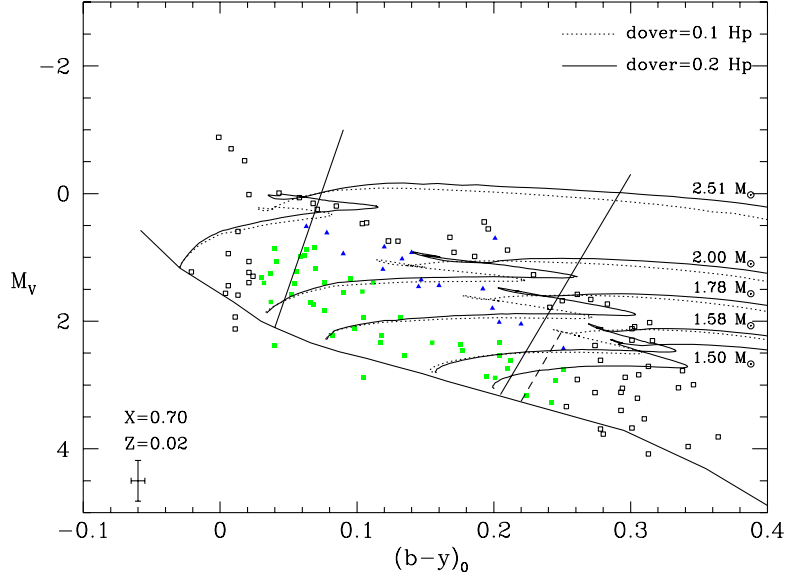}{6.5cm}{-90}{40}{40}{-160}{225}
      \caption{{\it Hipparcos} unreddened $M_V$ against dereddened
               $(b-y)_0$ colour indices for potential COROT targets in
               the Center direction. Dotted and solid lines indicate
               models with values of overshooting of $d_{\rm
               over}=0.1$ and $d_{\rm over}=0.2$, respectively.  Solid
               squares represent stars surely unevolved, independently
               of the value of the overshooting used. Solid triangles
               represent stars within the zigzags of the two sets of
               models and open squares are stars too evolved or
               outside of the instability strips.  Mean photometric
               errors are given by the error bars in the lower-left
               corner of the plot.}
   \end{figure}
   \begin{figure}
      \plotfiddle{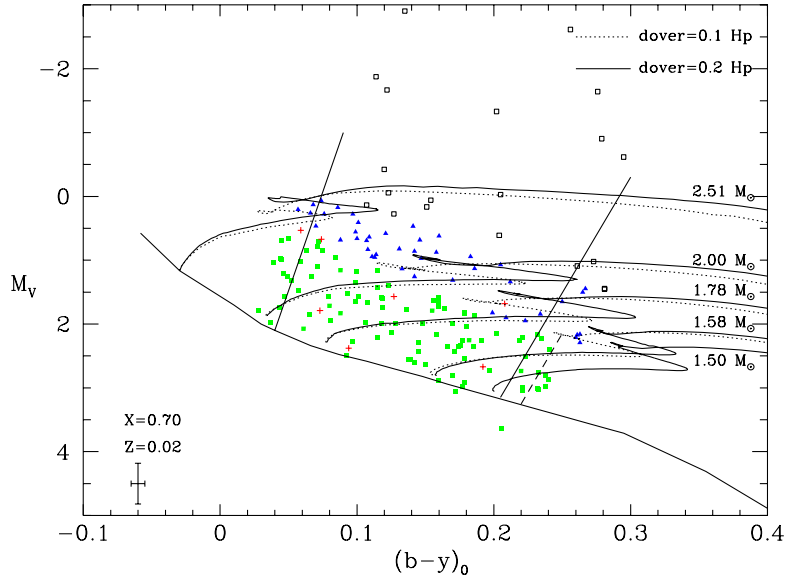}{6.5cm}{-90}{40}{40}{-160}{225}
      \caption{Same as in Fig.~1 but for the Anticenter direction.
               The crosses are star with no {\it Hipparcos} parallax
               and therefore with $M_V$ determined from the
               photometry.}
   \end{figure}

   \begin{figure}[t]
      \plotfiddle{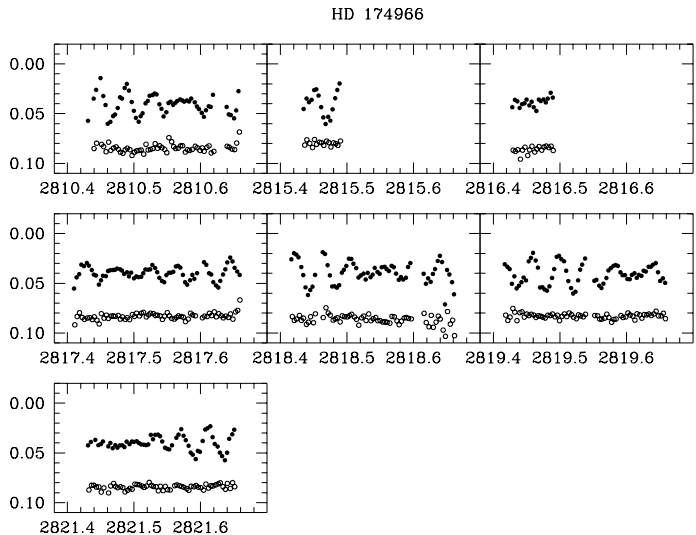}{6.5cm}{0}{50}{50}{-160}{-160}
      \caption{Str\"omgren $v$ differential (v$-$c) magnitude of the
      $\delta$~Sct star HD~174966 (solid circles) observed at OSN.
      Open circles are the differential light curve of the comparison
      stars HD~175272 and HD~175543.  The standard deviation of the
      measurements between the comparison stars is 0.0038 mag.}
   \end{figure}
   \begin{figure}
      \plotfiddle{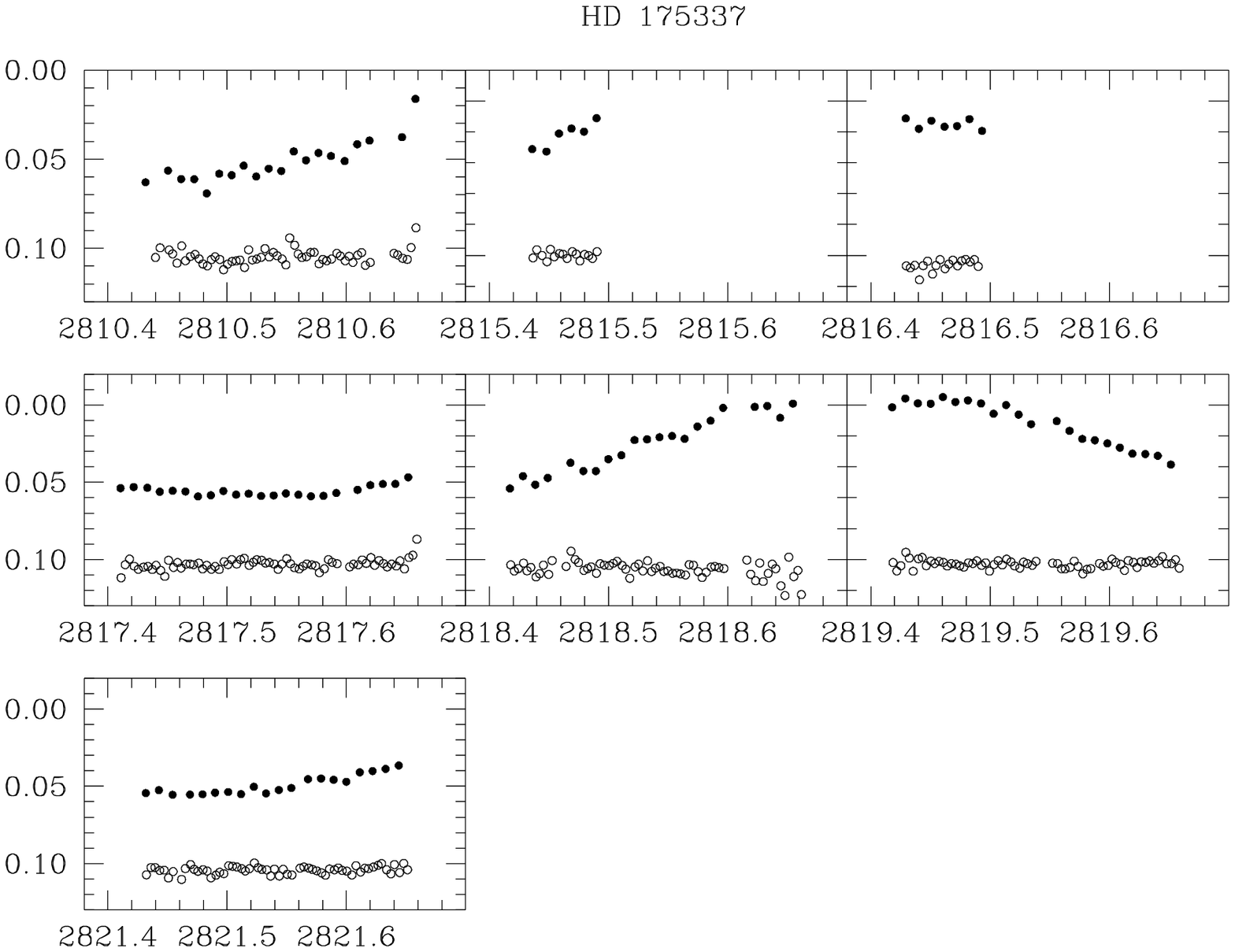}{6.5cm}{0}{50}{50}{-160}{-160}
      \caption{Same as Fig.~3 but for the $\gamma$~Dor star
      HD~175337}
   \end{figure}

\section{The observing program}

We have used the 1.5m and 0.9m telescopes of the Sierra Nevada
Observatory (OSN), managed by the IAA, to obtain $uvby\beta$ and
\ion{Ca}{ii} H\&K photometry of all the stars with $V \le 8.0$ mag
among which will be, to a large extent, the primary and secondary
COROT potential targets.  At the 1.5m telescope, \ion{Ca}{ii} H\&K CCD
photometry was gathered, while Str\"omgren and Crawford photoelectric
photometry was obtained simultaneously in the $uvby\beta$ filters at
the 0.9m telescope. The whole dataset was reduced and calibrated to
the respective standard systems by using specific software developed
by our group (Amado et al. in prep).  The GAUDI data server provides
access to COROT Archive of the Ground-Based Seismology Program and it
is developed and maintained by LAEFF, Spain (Solano et
al. 2003). These data has been utilised to determine $T_{\rm eff}$,
$\log{g}$ and [Fe/H] of all the possible targets of B, A and F
spectral types.

\section{Colour-Magnitude diagrams}

The observations mentioned above have, in part, led to the coverage of
the Colour-Magnitude diagram (CMD) on and near the main sequence in
the region of the Instability Strip (IS).  This has helped to enlarge
the sample of known $\delta$~Sct and $\gamma$~Dor stars. In Figs.~1
and 2, we present the diagrams for the stars within the IS in the
field of view of COROT (Center and Anticenter directions). The main
sequence is taken from Philip \& Egret (1980) and the $\delta$~Sct
instability strip borders from Rodr\'{\i}guez \& Breger (2001).  The
red border of the $\gamma$~Dor instability strip is taken from Handler
\& Shobbrook (2002); the blue border is well inside the $\delta$~Sct
instability strip.  We also added evolutionary tracks, considering
five values for the mass in the range from 1.50 to 2.51~$M_{\odot}$,
and hydrogen and metal content of 0.70 and 0.02, respectively.  We
calculated two sets of models for two different typical overshooting
extension distances, i.e., $d_{\rm over}=0.1$ and $d_{\rm over}=0.2$
(see Claret 1995 for details).  The limits of the IS for stars to the
left of the $\delta$~Sct IS blue edge and right of the $\gamma$~Dor IS
red edge have been a little relaxed to take into account the errors in
the photometric parameters.

\section{Enlarging the sample of known $\delta$~Sct and
$\gamma$~Dor stars}

Once the potential targets to be monitored were selected from their
position in the CMD, dedicated observing programs have been carried
out to try to make clear variability detections at the 0.005 mag
level.  The sample of variable stars discovered in the Center
direction is quite representative of the phenomenology we meet in the
lower part of the instability strip.  Figure~3 shows an example of
$\delta$~Sct variability: beat phenomena are clearly visible. Figure~4
shows an example of the $\gamma$~Dor variability, much slower than the
previous one.  Both stars have been observed during the two last weeks
of June 2003.  The work is still going on as for some of the stars the
variability has not yet been characterized in a precise way.  For more
examples of newly $\delta$~Sct and $\gamma$~Dor stars discovered in
the COROT ground--based observation program see Poretti, Garrido,
Amado et al. (2003).

\end{document}